\documentclass[11pt,fleqn]{article}

\usepackage{amsmath,amssymb,amsthm,enumerate,cite}

\newcommand{\p}{\partial}

\newcommand{\CV}{\mathop{\rm CV}\nolimits}
\newcommand{\CL}{\mathop{\rm CL}\nolimits}
\newcommand{\Ch}{\mathop{\rm Ch}\nolimits}

\newcounter{tbn}

\newcounter{mcasenum}

\newtheorem{theorem}{Theorem}

\newtheorem{corollary}{Corollary}
\newtheorem{proposition}{Proposition}
\newtheorem*{proposition*}{Proposition}
{\theoremstyle{definition}
\newtheorem{definition}{Definition}

\newtheorem{note}{Note}
}

\begin{document}

\par\noindent {\LARGE\bf
Equivalence of conservation laws\\ and equivalence of potential systems
\par}

{\vspace{4mm}\par\noindent {\bf N.M. Ivanova\,$^\dag$ and R.O. Popovych\,$^\ddag$
} \par\vspace{2mm}\par}
{\vspace{2mm}\par\noindent {\it
$^\dag{}^\ddag$~Institute of Mathematics of NAS of Ukraine, 3 Tereshchenkivska Str., 01601 Kyiv, Ukraine\\
}}
{\noindent \vspace{2mm}{\it
$\phantom{^\dag{}^\ddag}$~e-mail: ivanova@imath.kiev.ua, rop@imath.kiev.ua
}\par}
{\par\noindent\vspace{2mm} {\it
$^\ddag$~Fakult\"at f\"ur Mathematik, Universit\"at Wien, Nordbergstra{\ss}e 15, A-1090 Wien, Austria
} \par}

{\vspace{5mm}\par\noindent\hspace*{8mm}\parbox{140mm}{\small
We study conservation laws and potential symmetries of (systems of) differential equations applying equivalence relations
generated by point transformations between the equations.
A Fokker--Planck equation and the Burgers equation are considered as examples.
Using reducibility of them to the one-dimensional linear heat equation,
we construct complete hierarchies of local and potential conservation laws for them and describe, in some sense,
all their potential symmetries.
Known results on the subject are interpreted in the proposed framework.
This paper is an extended comment on the paper of J.-q.~Mei and H.-q.~Zhang [Internat. J. Theoret. Phys., 2006, in press].
}\par\vspace{5mm}}

\section{Introduction}

A number of authors investigate symmetries (in particular, potential ones) and conservation laws of different classes of equivalent equations.
In such way they perform a huge number of unnecessary cumbersome calculations.
Since equivalence transformations may be quite complicated,
sometimes it seems to be impossible to obtain directly complete and correct results for some cases which are equivalent to simple ones,
although these results can be easily reconstructed with application of equivalence transformations.

At the same time, it is a general mathematical rule that {\em equivalent in some sense objects possess equivalent properties}.
In particular, if two systems of equations are equivalent with respect to point transformations then there exists a one-to-one correspondence between
their maximal Lie invariance algebras, spaces of conservation laws, potential symmetries, exact solutions etc.
All the above mentioned features for more complicated models can be constructed from ones of the simpler equivalent models
by means of application of known equivalence transformation easier than by direct calculations.

In the presented paper we apply this observation to describe exhaustively local and potential conservation laws and potential symmetries
of a Fokker--Planck equation and the famous Burgers equation due to reducibility of them to the linear heat equation.

This paper was started as a comment on the recent work~\cite{Mei&Zhang2006}, where the authors tried to obtain some potential conservation laws
of the Burgers and Fokker--Planck equations, referring to results of~\cite{Pucci&Saccomandi1993,Saccomandi1997}.
At first our aim was to correct mathematically and historically inaccurate statements of~\cite{Mei&Zhang2006}.
Thus, e.g., system~(13) $v_x=u$, $v_t=u_x+xu$
is not a potential system for equation~(12) $u_t=(xu_x)_x+u_{xx}$ of~\cite{Mei&Zhang2006}.
Comparing~(12) and~(13) with~\cite{Pucci&Saccomandi1993,Saccomandi1997}
we deduce that the authors consider conservation laws of the Fokker--Planck equation of form $u_t=u_{xx}+(xu)_x$.
Let us mention also that the potential operator~(14) given in~\cite{Mei&Zhang2006}
with a reference to~\cite{Pucci&Saccomandi1993,Saccomandi1997} is written with missprints (up to the multiplier $e^{2t}$).
Therefore, the conservation law obtained with this operator may be incorrect.
Potential symmetry operators for the Burgers equation obtained in~\cite{Mei&Zhang2006} do not form linear
space and, therefore, do not form a Lie algebra.
Direct calculations show also that $(T^1,T^2)$ adduced in the last formula of the paper~\cite{Mei&Zhang2006}
is not a conserved vector for the Burgers equation.
While working on the comment we decided to generalize essentially results of~\cite{Mei&Zhang2006}
adducing also the complete sets of the potential conservation laws
and giving a complete, in some sense, description of all possible potential symmetries of the Burgers and Fokker--Planck equations.

To construct non-local conservation laws of the Fokker--Planck and Burgers equations
the following statement~\cite{Ibragimov&Kara&Mahomed1998,Kara&Mahomed2000} was used in~\cite{Mei&Zhang2006}
(see also~\cite{Khamitova1982,Ibragimov1985,Popovych&Ivanova2004ConsLawsLanl}).
\begin{proposition}
If the system~$\mathcal{L}$ admits a one-parameter group of symmetry transformations with the infinitesimal generator
$Q=\tau\p_t+\xi\p_x+\eta\p_{u}$ and a conservation law of form $D_tT+D_xX=0$, then the vector-function~$(\tilde T,\tilde X)$ with
\begin{gather*} 
\tilde T=Q(T)+TD_x\xi-XD_x\tau,\quad \tilde X=Q(X)+XD_t\tau-TD_t\xi
\end{gather*}
is a conserved vector.
\end{proposition}
If the conservation law is invariant with respect to the action of operator~$Q$ then~$(\tilde T,\tilde X)$ vanishes.
Thus, one obtains a rule for finding conservation laws invariant with respect to symmetry operator:
\[
Q(T)+TD_x\xi-XD_x\tau=0,\quad Q(X)+XD_t\tau-TD_t\xi=0.
\]
Although the latter formula (Theorem~3.1 from~\cite{Mei&Zhang2006}) is completely true
and can be useful for obtaining particular classes of conservation laws,
it is not guarantee that the resulting conservation law will be non-trivial or pure nonlocal.
Thus, e.g., if one redo the calculations leading to formula~(26) of~\cite{Mei&Zhang2006}
without computational inaccuracies, one obtains exactly conservation law equivalent to local one.
In general, invariant conservation laws do not generate the complete space of conservation laws.
Therefore, this approach is suitable only in case the complete set of conservation laws is difficult or even impossible to be constructed.
As shown in~\cite{Popovych&Ivanova2004ConsLawsLanl}, complete hierarchies of potential conservation laws
for both equations under consideration can be constructed.

Below we apply the equivalence framework to examine potential symmetries and conservation laws of the Fokker--Planck and Burgers equations
and to give a detailed interpretation of interesting results of Pucci and Saccomandi~\cite{Pucci&Saccomandi1993,Saccomandi1997}.
Since the both equations under consideration are reduced to the linear heat equation,
we also describe a way to find potential symmetries of the linear heat equation using potential systems associated with nonconstant characteristics.

More precisely, our paper is organized as follows.
At first, in Section~\ref{SectionCLDef} we adduce a necessary theoretical background on
conservation laws and potential systems, including a short discussion of empiric and rigorous definitions of conservation laws.
Equivalences of conservation laws and potential systems with respect to point transformations are also discussed in detail.
Then, using reducibility of the Fokker--Planck and Burgers equations to the one-dimensional linear heat equation,
we construct complete hierarchies of local and potential conservation laws for them and describe
their potential symmetries (Sections~\ref{SectionFokPlEq} and~\ref{SectionBurgersEq}).
Potential symmetries of the linear heat equation are considered in Section~\ref{SectionOnPotSymmetriesOfLHE}.

\section{Theoretical background}\label{SectionCLDef}

To begin with, we adduce a necessary theoretical background on conservation laws and potential symmetries,
following, e.g.,~\cite{Olver1986,Popovych&Ivanova2004ConsLawsLanl,Zharinov1986} and
considering for simplicity the case of two independent variables~$t$ (the time variable) and~$x$ (the space variable).
See the above references for the general case.

Let~$\mathcal{L}$ be a system~$L(t,x,u_{(\rho)})=0$ of $l$ PDEs $L^1=0$, \ldots, $L^l=0$
for the unknown functions $u=(u^1,\ldots,u^m)$
of the independent variables~$t$ and~$x$.
Here $u_{(\rho)}$ denotes the set of all the partial derivatives of the functions $u$
of order not greater than~$\rho$, including $u$ as the derivatives of the zero order.

First we give an empiric definition of conservation laws.

\begin{definition}\label{def.conservation.law}
A {\em conservation law} of the system~$\mathcal{L}$ is a divergence expression
\begin{equation}\label{conslaw}
D_tT(t,x,u_{(r)})+D_xX(t,x,u_{(r)})=0
\end{equation}
which vanishes for all solutions of~$\mathcal{L}$.
Here $D_t$ and $D_x$ are the operators of total differentiation with respect to $t$ and $x$, respectively.
The differential functions $T$ and $X$ are correspondingly called a {\em density} and a {\em flux} of the conservation law and
the tuple $(T,X)$ is a \emph{conserved vector} of the conservation law.
\end{definition}

The crucial notion of the theory of conservation laws is one of equivalence and triviality of conservation laws.
\begin{definition}\label{DefinitionOfConsVectorEquivalence}
Two conserved vectors $(T,X)$ and $(T',X')$ are {\em equivalent} if
there exist functions~$\hat T$, $\hat X$ and~$H$ of~$t$, $x$ and derivatives of~$u$ such that
$\hat T$ and $\hat X$ vanish for all solutions of~$\mathcal{L}$~and
$T'=T+\hat T+D_xH$, $X'=X+\hat X-D_tH$.
A conserved vector is called {\em trivial} if it is equivalent to the zeroth vector.
\end{definition}

The notion of linear dependence of conserved vectors is introduced in a similar way.
Namely, a set of conserved vectors is {\em linearly dependent}
if a linear combination of them is a trivial conserved vector.

It is obvious that under the problem of finding conservation laws for some system one should
understand the problem of finding {\em inequivalent linearly independent} conservation laws, i.e.,
conservation laws having linearly independent conserved vectors.

Conservation laws can be investigated in the above empiric framework.
However, for deeper understanding of the problem and absolutely correct calculations
a more rigorous definition of conservation laws should be used.

For any system~$\mathcal{L}$ of differential equations the set~$\CV(\mathcal{L})$ of conserved vectors of
its conservation laws is a linear space,
and the subset~$\CV_0(\mathcal{L})$ of trivial conserved vectors is a linear subspace in~$\CV(\mathcal{L})$.
The factor space~$\CL(\mathcal{L})=\CV(\mathcal{L})/\CV_0(\mathcal{L})$
coincides with the set of equivalence classes of~$\CV(\mathcal{L})$ with respect to the equivalence relation adduced in
Definition~\ref{DefinitionOfConsVectorEquivalence}.

\begin{definition}\label{DefinitionOfConsLaws}
The elements of~$\CL(\mathcal{L})$ are called {\em conservation laws} of the system~$\mathcal{L}$,
and the whole factor space~$\CL(\mathcal{L})$ is called {\em the space of conservation laws} of~$\mathcal{L}$.
\end{definition}

That is why description of the set of conservation laws can be assumed
as finding~$\CL(\mathcal{L})$ that is equivalent to construction of either a basis if
$\dim \CL(\mathcal{L})<\infty$ or a system of generatrices in the infinite dimensional case.
The elements of~$\CV(\mathcal{L})$ which belong to the same equivalence class giving a conservation law~${\cal F}$
are considered all as conserved vectors of this conservation law,
and we will additionally identify elements from~$\CL(\mathcal{L})$ with their representatives
in~$\CV(\mathcal{L})$.
For $(T,X)\in\CV(\mathcal{L})$ and ${\cal F}\in\CL(\mathcal{L})$
the notation~$(T,X)\in {\cal F}$ will denote that $(T,X)$ is a conserved vector corresponding
to the conservation law~${\cal F}$.
In contrast to the order $r_{(T,X)}$ of a conserved vector~$(T,X)$ as the maximal order of derivatives explicitly appearing in
the differential functions $T$ and~$X$,
the {\em order of the conservation law}~$\cal F$
is called $\min\{r_{(T,X)}\,|\,(T,X)\in{\cal F}\}$.
Under linear dependence of conservation laws we understand linear dependence of them as elements of~$\CL(\mathcal{L})$.
Therefore, in the framework of ``representative'' approach
conservation laws of a system~$\mathcal{L}$ are considered as {\em linearly dependent} if
there exists linear combination of their representatives, which is a trivial conserved vector.

Let the system~$\cal L$ be totally nondegenerate~\cite{Olver1986}.
Then application of the Hadamard lemma to the definition of conservation law and integrating by parts imply that
the left hand side of any conservation law of~$\mathcal L$ can be always presented up to the equivalence relation
as a linear combination of left hand sides of independent equations from $\mathcal L$
with coefficients~$\lambda^\mu$ being functions of $t$, $x$ and derivatives of~$u$:
\begin{equation}\label{CharFormOfConsLaw}
D_tT+D_xX=\lambda^1 L^1+\dots+\lambda^l L^l.
\end{equation}

\begin{definition}\label{DefCharForm}
Formula~\eqref{CharFormOfConsLaw} and the $l$-tuple $\lambda=(\lambda^1,\ldots,\lambda^l)$
are called the {\it characteristic form} and the {\it characteristic}
of the conservation law~$D_tT+D_xX=0$ correspondingly.
\end{definition}

The characteristic~$\lambda$ is {\em trivial} if it vanishes for all solutions of $\cal L$.
Since $\cal L$ is nondegenerate, the characteristics~$\lambda$ and~$\tilde\lambda$ satisfy~\eqref{CharFormOfConsLaw}
for the same conserved vector~$(T,X)$ and, therefore, are called {\em equivalent}
iff $\lambda-\tilde\lambda$ is a trivial characteristic.
Similarly to conserved vectors, the set~$\Ch(\mathcal{L})$ of characteristics
corresponding to conservation laws of the system~$\cal L$ is a linear space,
and the subset~$\Ch_0(\mathcal{L})$ of trivial characteristics is a linear subspace in~$\Ch(\mathcal{L})$.
The factor space~$\Ch_{\rm f}(\mathcal{L})=\Ch(\mathcal{L})/\Ch_0(\mathcal{L})$
coincides with the set of equivalence classes of~$\Ch(\mathcal{L})$
with respect to the above characteristic equivalence relation.

Any conservation law~\eqref{conslaw} of~$\mathcal{L}$ allows us to deduce the new dependent (potential) variable~$v$
by means of the equations
\begin{equation}\label{potsys1}
v_x=T,\qquad v_t=-X.
\end{equation}

In the case of single equation~$\mathcal{L}$, equations of form~\eqref{potsys1} combine into
the complete potential system since~$\mathcal{L}$ is a differential consequence of~\eqref{potsys1}.
As a rule, systems of such kind admit a number of nontrivial symmetries and so they are of a great interest.
If the transformation of some of nonlocal variables~$t$, $x$ or~$u$ depends explicitly on
variable~$v$, such symmetry is a nonlocal for the initial equation (system) and is called
{\em potential symmetry}.
Let us mention that the concept of potential symmetry was introduced by
Bluman {\it at al}~\cite{Bluman1989,Bluman&Reid&Kumei1988} in the late 80-es.
See also the related notion of quasilocal symmetry~\cite{Akhatov&Gazizov&Ibragimov1987,Akhatov&Gazizov&Ibragimov1989}.

An important property of the class of equations in the conserved form is that it is preserved
under any point transformation (see, e.g.,~\cite{Popovych&Ivanova2004ConsLawsLanl}).

\begin{proposition}\label{PropEqTrCL}
A point transformation~$g$: $\tilde t=t^g(t,x,u)$, $\tilde x=x^g(t,x,u)$, $\tilde u=u^g(t,x,u)$ prolonged to derivatives of~$u$
transforms the equation $D_tT+D_xX=0$ to the equation \mbox{$D_tT^g+D_xX^g=0$}. The transformed conserved vector~$(T^g,X^g)$ is determined
by the formula
\begin{gather*}
T^g(\tilde x,\tilde u_{(r)})=\frac{T(x,u_{(r)})D_t\tilde t+X(x,u_{(r)})D_x\tilde t}{D_t\tilde t\,D_x\tilde x-D_x\tilde t\,D_t\tilde x},
\\[1ex]
X^g(\tilde x,\tilde u_{(r)})=\frac{T(x,u_{(r)})D_t\tilde x+X(x,u_{(r)})D_x\tilde x}{D_t\tilde t\,D_x\tilde x-D_x\tilde t\,D_t\tilde x}.
\end{gather*}
\end{proposition}

\begin{note}
In the case of one dependent variable ($m=1$) $g$ can be a contact transformation:
$\tilde t=t^g(t,x,u_{(1)})$, $\tilde x=x^g(t,x,u_{(1)})$, $\tilde u_{(1)}=u^g_{(1)}(t,x,u_{(1)})$.
Similar notes are also true for the below statements.
\end{note}

\begin{proposition}\label{PropositionOnInducedMapping}
Any point transformation $g$ between systems~$\mathcal{L}$ and~$\tilde{\mathcal{L}}$
induces a linear one-to-one mapping $g_*$ from~$\CV(\mathcal{L})$ into~$\CV(\tilde{\mathcal{L}})$,
which maps $\CV_0(\mathcal{L})$ into~$\CV_0(\tilde{\mathcal{L}})$
and generates a linear one-to-one mapping $g_{\rm f}$ from~$\CL(\mathcal{L})$ into~$\CL(\tilde{\mathcal{L}})$.
\end{proposition}

\begin{corollary}\label{CorollaryOnInducedMappingOfChar}
Any point transformation $g$ between systems~$\mathcal{L}$ and~$\tilde{\mathcal{L}}$
induces a linear one-to-one mapping $\hat g_{\rm f}$ from~$\Ch_{\rm f}(\mathcal{L})$
into~$\Ch_{\rm f}(\tilde{\mathcal{L}})$.
\end{corollary}

It is possible to obtain an explicit formula for correspondence between characteristics of~$\mathcal{L}$
and~$\tilde{\mathcal{L}}$.
Let $\tilde{\mathcal{L}}^\mu=\Lambda^{\mu\nu}\mathcal{L}^\nu$,
where $\Lambda^{\mu\nu}=\Lambda^{\mu\nu\alpha}D^\alpha$, $\Lambda^{\mu\nu\alpha}$ are differential functions,
$\alpha=(\alpha_t,\alpha_x)$ runs the multi-indices set ($\alpha_i\!\in\!\mathbb{N}\cup\{0\}$),
$\mu,\nu=\overline{1,l}$.
Then
\[\lambda^\mu={\Lambda^{\nu\mu}}^*((D_t\tilde t\,D_x\tilde x-D_x\tilde t\,D_t\tilde x)\tilde\lambda^\nu).\]
Here ${\Lambda^{\nu\mu}}^*=(-D)^\alpha\cdot\Lambda^{\mu\nu\alpha}$ is the adjoint to the operator~$\Lambda^{\nu\mu}$.
For a number of cases, e.g. if~$\mathcal{L}$ and~$\tilde{\mathcal{L}}$ are single partial differential equations
($l=1$), the operators~$\Lambda^{\mu\nu}$ are simply differential functions
(i.e., $\Lambda^{\mu\nu\alpha}=0$ for $|\alpha|>0$) and, therefore, ${\Lambda^{\nu\mu}}^*=\Lambda^{\mu\nu}$.

Equivalent conservation laws give rise to equivalent potential systems.
More precisely, Proposition~\ref{PropEqTrCL} and equations~\eqref{potsys1} imply the following statement.

\begin{proposition}\label{2DConsLawEquivRelation}
Any point transformation connecting two systems~$\mathcal{L}$ and~$\tilde{\mathcal L}$
of PDEs with two independent variables generates a one-to-one mapping between the sets of potential systems,
which correspond to~$\mathcal{L}$ and~$\tilde{\mathcal L}$. Generation is made via trivial prolongation
on the space of introduced potential variables, i.e., we can assume that the potentials are not transformed.
\end{proposition}


In such way, if a transformation connects two systems of differential equations, then the same transformation
maps the set of conservation laws  of the first system to the set of conservation laws of the second system
and the space of characteristics of conservation laws to the space of characteristics.
This transformation is trivially prolonged to the possible potential variables and then
makes a mapping between the corresponding sets of potential systems.
Therefore, one can easily derive characteristics of conservation laws, conservation laws,
potential systems and potential symmetries of the more complicated system from the ones of the simpler system.

\section{Fokker--Planck equation}\label{SectionFokPlEq}

In~\cite{Pucci&Saccomandi1993,Saccomandi1997} Pucci and Saccomandi investigated potential symmetries of the Fokker--Planck equation
\begin{equation}\label{eqFokPlEq}
u_t=u_{xx}+(xu)_x.
\end{equation}
Using the simplest conservation law $D_tu-D_x(u_x+xu)=0$ with the characteristic~$1$, they construct
the corresponding potential system
\begin{equation}\label{sysPotSysFocPlEqChar1}
v_x=u,\quad v_t=u_x+xu
\end{equation}
and then found its maximal Lie invariance algebra
\begin{gather*}
\mathfrak g_1=\langle\p_t,\ e^{-t}\p_x,\ e^{-2t}\p_t-e^{-2t}x\p_x+e^{-2t}u\p_u,\ e^t\p_x-e^t(xu+v)\p_u-e^txv\p_v,\\
\qquad\ e^{2t}\p_t+e^{2t}x\p_x-e^{2t}(x^2u+2xv+2u)\p_u-e^{2t}(x^2+1)v\p_v,\ u\p_u+v\p_v,\ f_x\p_u+f\p_v \rangle,
\end{gather*}
where the function $f=f(t,x)$ runs the solution set of the equation $f_t=f_{xx}+xf_x$.
Any operator from $\mathfrak g_1$ is a potential symmetry operator of equation~\eqref{eqFokPlEq}.
It is a nontrivial potential symmetry operator iff the coefficient of $\p_u$ depends on~$v$.
Let us note that the algebra $\mathfrak g_1$ of potential symmetry operators differs from the maximal Lie invariance algebra
\begin{gather*}
\mathfrak g_0=\langle\p_t,\ e^{-t}\p_x,\,e^{-2t}\p_t-e^{-2t}x\p_x+e^{-2t}u\p_u,\ e^t\p_x-e^txu\p_u,\\
\qquad\ e^{2t}\p_t+e^{2t}x\p_x-e^{2t}x^2u\p_u,\ u\p_u,\ f_x\p_u+f\p_v \rangle
\end{gather*}
of equation~\eqref{eqFokPlEq} and is not projectible to $\mathfrak g_0$.
At the same time, the algebras $\mathfrak g_0$ and $\mathfrak g_1$ are isomorphic.

We interpret the results by Pucci and Saccomandi on the characteristic~$1$ in the above framework of the equivalence relation between
conservation laws of different equations, which is extended to equivalence of potential systems.
Then we give a generalization for the case of arbitrary characteristic.

It is well-known that the  Fokker--Planck equation~\eqref{eqFokPlEq} is reduced to the linear heat equation
$\tilde u_{\tilde t}=\tilde u_{\tilde x\tilde x}$ by the transformation
\begin{gather*}
\mathcal T\colon\quad\tilde t=\frac12e^{2t},\quad \tilde x=e^tx,\quad \tilde u=e^{-t}u
.
\end{gather*}
The same transformation maps the conservation law of the Fokker--Planck equation with characteristic~$1$
to the one of the linear heat equation with the same characteristic.
In view of Proposition~\ref{2DConsLawEquivRelation}, potential system~\eqref{sysPotSysFocPlEqChar1} reduces to the potential system
\begin{gather}
\label{sysPotSysLHEChar1}
\tilde v_{\tilde x}=\tilde u,\quad \tilde v_{\tilde t}=\tilde u_{\tilde x}
\end{gather}
of the linear heat equation by the transformation~$\mathcal T$ trivially prolonged to the potential variable $\tilde v=v$.
Therefore, the prolonged transformation~$\mathcal T_\mathrm{pr}$ establishes an isomorphism between
the algebra~$\mathfrak g_1$ and maximal Lie invariance algebra
\begin{gather*}
\tilde{\mathfrak g}_1=\langle
\p_t,\ \p_x,\ 2t\p_t+x\p_x-u\p_u,\ 2t\p_x-(xu+v)\p_u-xv\p_v, \\
\qquad\ 4t^2\p_t+4tx\p_x-((x^2+6t)u+2xv)\p_u-(x^2+2t)v\p_v,\ u\p_u+v\p_v,\ h_x\p_u+h\p_v
\rangle,
\end{gather*}
of system~\eqref{sysPotSysLHEChar1} in a similar way as the simple transformation~$\mathcal T$ do this between
the algebra~$\mathfrak g_0$ and maximal Lie invariance algebra
\begin{gather*}
\tilde{\mathfrak g}_0=\langle
\p_t,\ \p_x,\ 2t\p_t+x\p_x,\ 2t\p_x-xu\p_u,\ 4t^2\p_t+4tx\p_x-(x^2+2t)u\p_u,\ u\p_u,\ h\p_u\rangle
\end{gather*}
of the linear heat equation.
Here the function $h=h(t,x)$ runs the solution set of this equation. Tildes over variables in the formulas for
$\tilde{\mathfrak g}_0$ and $\tilde{\mathfrak g}_1$ are omitted.

As shown in~\cite{Popovych&Ivanova2003PETs}, the algebra~$\tilde{\mathfrak g}_1$ also is isomorphic
to the maximal Lie invariance algebra of the potential equation $\tilde v_{\tilde t}=\tilde v_{\tilde x\tilde x}$
(the form of which coincides with the initial equation on~$u$).
Moreover, the Lie symmetry operators of the potential system~\eqref{sysPotSysLHEChar1} are the first prolongations
of the Lie symmetry operators of the potential equation.
It means that a similar statement is true for all equivalent equations.
Namely, the Lie symmetry operators of the potential system~\eqref{sysPotSysFocPlEqChar1}
 are the first prolongations of the Lie symmetry operators
of the corresponding potential equation $v_t=v_{xx}+xv_x$
that can be reduced to the ``potential'' linear heat equation $\tilde v_{\tilde t}=\tilde v_{\tilde x\tilde x}$
by means of the truncated transformation $\tilde t=\frac12e^{2t}$, $\tilde x=e^tx$, $\tilde v=v$.

The set of all possible linearly independent local conservation laws of the linear heat equation
is well-known~\cite{Dorodnitsyn&Svirshchevskii1983,Ivanova2004,Ivanova&Popovych&Sophocleous2005,Olver1986,
Popovych&Ivanova2004ConsLawsLanl,Steinberg&Wolf1981}
and consists of ones having the form
\[
D_{\tilde t}(\tilde\alpha\tilde u)+D_{\tilde x}(\tilde\alpha_{\tilde x}\tilde u-\tilde\alpha \tilde u_{\tilde x})=0,
\]
where $\tilde\alpha=\tilde\alpha(\tilde t,\tilde x)$ is an arbitrary solution of the backward linear heat equation
\mbox{$\tilde\alpha_{\tilde t}+\tilde\alpha_{\tilde x\tilde x}=0$}.
Using Proposition~\ref{PropEqTrCL} or different versions of the direct method for
finding conservation laws~\cite{Anco&Bluman2002a,Anco&Bluman2002b,Popovych&Ivanova2004ConsLawsLanl,Wolf2002},
we obtain the set of linearly independent local conservation laws of the Fokker--Planck equation~\eqref{eqFokPlEq}
\begin{equation}\label{CLlocCLFPE}
D_t(\alpha u)+D_x((\alpha_x-x\alpha)u-\alpha u_x)=0,
\end{equation}
where $\alpha=\alpha(t,x)$ is an arbitrary solution of the linear equation
$\alpha_t+\alpha_{xx}-x\alpha_x=0$, which is adjoint to the Fokker--Planck equation.

In~\cite{Popovych&Ivanova2004ConsLawsLanl} the theorem was proved that any
potential conservation law of the linear heat equation is equivalent to the local one.
Therefore, in view of Proposition~\ref{PropEqTrCL} the same is true for the Fokker--Planck equation~\eqref{eqFokPlEq},
and formula~\eqref{CLlocCLFPE} gives the complete description of the local and potential conservation laws
of the Fokker--Planck equation~\eqref{eqFokPlEq}.

Let us emphasize that the above statement on the potential conservation laws is true not only for the laws
obtained from the potential system corresponding to the characteristic~$\alpha=1$, but also
for the system which is a union of any finite number of the potential systems corresponding to linearly
independent solutions of the backward heat equation.

\section{On potential symmetries of linear heat equation}\label{SectionOnPotSymmetriesOfLHE}

In contrast to the conservation laws, potential symmetries of the linear heat equation are investigated only in case of the single characteristics~$1$.
The problem of construction of all possible potential symmetries of the linear heat equation remains open
(in particular, it includes investigation of systems with arbitrary families of linearly independent characteristics).

Here we study the families of potential systems of the linear heat equation which are constructed with single local conservation laws
and make a preliminary investigation.
As mentioned in Section~\ref{SectionFokPlEq}, any local conservation law of the linear heat equation $u_t=u_{xx}$
has the form
\[
D_t(\alpha u)+D_x(\alpha_xu-\alpha u_x)=0,
\]
where $\alpha=\alpha(t,x)$ is an arbitrary non-zero solution of the backward heat equation $\alpha_t+\alpha_{xx}=0$.
The characteristic of the conservation law coincides with $\alpha$.
The associated potential system is
\begin{equation}\label{EqGenSimplestPotSystem}
v_x=\alpha u, \quad v_t=\alpha u_x-\alpha_xu.
\end{equation}
The initial equation on $u$ is a differential consequence of system~\eqref{EqGenSimplestPotSystem}.
Another differential consequence is the equation
\begin{equation}\label{EqGenSimplestPotEq}
v_t+2\frac{\alpha_x}\alpha v_x-v_{xx}=0
\end{equation}
on the potential dependent variable~$v$ which is called the potential equation associated with the equation~$u_t=u_{xx}$
and the characteristic~$\alpha$.

Consider a Lie symmetry operator $Q=\tau\p_t+\xi\p_x+\eta\p_u+\theta\p_v$ of system~\eqref{EqGenSimplestPotSystem}.
The coefficients of~$Q$ are functions of $t$, $x$, $u$ and $v$.
The infinitesimal invariance criterion~\cite{Olver1986,Ovsiannikov1982} implies for system~\eqref{EqGenSimplestPotSystem} that, in particular,
\begin{gather}
\tau_u=\xi_u=\theta_u=0,\label{EqTuple1OfDetEqsForLieSymmetriesOfGenSimplestPotSystem}\\[.5ex]
\tau_x=\tau_v=\xi_v=\theta_{vv}=0,\label{EqTuple2OfDetEqsForLieSymmetriesOfGenSimplestPotSystem}\\
\eta=\left(\theta_v-\xi_x-\frac{\alpha_t}\alpha\tau-\frac{\alpha_x}\alpha\xi\right)u+\frac{\theta_x}\alpha.
\label{EqTuple3OfDetEqsForLieSymmetriesOfGenSimplestPotSystem}
\end{gather}
The subsystem~\eqref{EqTuple1OfDetEqsForLieSymmetriesOfGenSimplestPotSystem} of determining equations means that
any Lie symmetry transformation of~\eqref{EqGenSimplestPotSystem} with respect to $t$, $x$ and $v$ does not depend on~$u$.
Equation~\eqref{EqGenSimplestPotEq} is a differential consequence of system~\eqref{EqGenSimplestPotSystem},
and there exist one-to-one correspondence between the sets of solutions of equation~\eqref{EqGenSimplestPotEq} and system~\eqref{EqGenSimplestPotSystem}.
Therefore, the truncated operator $\hat Q=\tau\p_t+\xi\p_x+\theta\p_v$ is a Lie symmetry operator of equation~\eqref{EqGenSimplestPotEq}.

And vice versa, consider a Lie symmetry operator $\hat Q=\tau\p_t+\xi\p_x+\theta\p_v$ of equation~\eqref{EqGenSimplestPotEq}.
The coefficients of~$\hat Q$ are functions of $t$, $x$ and $v$.
Then the prolonged to~$u$ operator $Q=\hat Q+\eta\p_u$, where $\eta$ is defined
by formula~\eqref{EqTuple3OfDetEqsForLieSymmetriesOfGenSimplestPotSystem}, is
a Lie symmetry operator of system~\eqref{EqGenSimplestPotSystem}.

In view of the subsystem~\eqref{EqTuple2OfDetEqsForLieSymmetriesOfGenSimplestPotSystem} of determining equations and
the formula for~$\eta$, we have \mbox{$\eta_v=\theta_{xv}$}.
Hence the conservation law with the characteristic~$\alpha$ results to pure potential symmetries of the linear heat equation
iff the potential equation~\eqref{EqGenSimplestPotEq} possesses a Lie symmetry operator with the coefficient~$\theta$ of~$\p_v$,
which satisfy the condition $\theta_{xv}\ne0$.

Let us study the case of the simplest nonconstant characteristic $\alpha=x$.
The corresponding potential equation has the form
\begin{equation}\label{EqGenSimplestPotEqCharx}
v_t+\frac2x v_x-v_{xx}=0.
\end{equation}
The maximal Lie invariance algebra of~\eqref{EqGenSimplestPotEqCharx} is
\[
\mathfrak t_0=\langle
\p_t,\ 2t\p_t+x\p_x,\ 4t^2\p_t+4tx\p_x-(x^2-2t)v\p_v,\ v\p_v,\ f\p_v\rangle,
\]
where the function $f=f(t,x)$ runs the solution set of~\eqref{EqGenSimplestPotEqCharx}.
It is easy to see that for the third basis operator the condition $\theta_{xv}\ne0$ is satisfied.
Therefore, consideration with the characteristic $\alpha=x$ leads to pure potential symmetries of the linear heat equation.
Namely, the system~\eqref{EqGenSimplestPotSystem} with $\alpha=x$ possesses the maximal Lie invariance algebra
\begin{gather*}
\mathfrak t_1=\langle
\p_t,\ 2t\p_t+x\p_x-2u\p_u,\ 4t^2\p_t+4tx\p_x-((x^2+6t)u+2v)\p_u-(x^2-2t)v\p_v,\\
\qquad\ u\p_u+v\p_v,\ x^{-1}f_x\p_u+f\p_v\rangle,
\end{gather*}
where the function $f=f(t,x)$ again runs the solution set of~\eqref{EqGenSimplestPotEqCharx}.
Any linear combination of operators from $\mathfrak t_1$ which contains the third basis operator is
an pure potential symmetry operator of the linear heat equation.
Note that the pure potential symmetry operators from the algebra~$\mathfrak t_1$ differ from ones from the algebra~$\mathfrak g_1$
both the explicit form and the nature of the potential variable which is defined by system~\eqref{EqGenSimplestPotSystem} with $\alpha=x$
instead of system~\eqref{sysPotSysLHEChar1}.

\section{Burgers equation}\label{SectionBurgersEq}

The famous Burgers equation
\begin{equation}\label{eqBurgersEq}
u_t=u_{xx}+2uu_x
\end{equation}
is singular among nonlinear evolution equations due to its symmetry properties
and admits the five-dimensional maximal Lie invariance algebra~\cite{Olver1986}
\begin{gather*}
\mathfrak b_0=\langle
\p_t,\ \p_x,\ t\p_x-\p_u,\ 2t\p_t+x\p_x-u\p_u,\ t^2\p_t+tx\p_x-(tu+x)\p_u\rangle.
\end{gather*}
It has one linearly independent local conservation law $D_tu-D_x(u_x+u^2)=0$, the characteristic of which equals to 1.
The Lie symmetries of the potential system
\[
v_x=u,\quad v_t=u_x+u^2
\]
associated with this conservation law
are also well-known and studied by many authors. See, e.g.,~\cite{Sophocleous1996,Popovych&Ivanova2003PETs}.
Its maximal Lie invariance algebra
\begin{gather*}
\mathfrak b_1=\langle\p_t,\ \p_x,\,\p_v,\ 2t\p_t+x\p_x-u\p_u,\ 2t\p_x-\p_u-2x\p_v,\\ \qquad\
4t^2\p_t+4tx\p_x-2(x+2tu)\p_u-(x^2+2t)\p_v,\ e^{-v}(\alpha_x-\alpha u)\p_u-4e^{-v}\alpha\p_v \rangle,
\end{gather*}
is infinite-dimensional. Here $\alpha=\alpha(t,x)$ is an arbitrary solution of the linear heat equation $\alpha_{t}=\alpha_{xx}$.
The elements of~$\mathfrak b_1$ are the first prolongations of Lie symmetry operators of the potential Burgers equation
$v_t=v_{xx}+v_x^2$ that is equivalent to the linear heat equation under the transformation
$\tilde t=t$, $\tilde x=x$, $\tilde v=e^v$.
Since this transformation is not point in the variables of the initial equation ($\tilde u=ue^v$),
the structure of the space of potential conservation laws of the Burgers equation
is more complicated than in case of the linear heat equation.
Namely, the following theorem is proved in~\cite{Popovych&Ivanova2004ConsLawsLanl}.

\begin{theorem}
The complete hierarchy of the potential conservation
laws of the Burgers equation consists of one local conservation law \[D_tu-D_x(u_x+u^2)=0\] and infinite number of the linearly independent
potential conservation laws have the form \[D_t(\beta e^v)+D_x((\beta_x-\beta u)e^v)=0\]
parameterized by the linearly independent solutions $\beta=\beta(t,x)$ of the backward linear heat equation $\beta_t=\beta_{xx}$.
\end{theorem}

The potential systems of the second level of the Burgers equation
(i.e., potential systems obtained with usage of potential conservation laws) are equivalent to potential systems of the linear heat equation.
That is why, investigation of the second-level potential symmetries is reduced to investigation of potential symmetries
of the linear heat equation.

\section{Conclusion}

In the presented paper we construct potential symmetries and complete hierarchies of potential conservation laws
of the Fokker--Planck and Burgers equations via reduction of them to the linear heat equation.
A brief discussion is given for the potential symmetries of the linear heat equation obtained from the potential system
associated with an arbitrary single conservation law of the linear heat equation. The case of characteristics~1 and~$x$ are studied exhaustively.

Let us emphasize that the problem on potential symmetries of the linear heat equation
have been solved only partially and is still open in the general statement.
This problem includes investigation of potential systems~\eqref{EqGenSimplestPotSystem} with arbitrary characteristics
and of union of such systems with arbitrary families of linearly independent characteristics.
In particular, it is necessary to find such conditions for families of characteristics that the associated potential system
yields pure potential symmetries for the linear heat equation.

The next important and interesting generalization of the obtained results
is to realize the same program for arbitrary $(1+1)$-dimensional linear parabolic equations.
We strongly believe that the method used in~\cite{Popovych&Ivanova2004ConsLawsLanl} for construction of all possible potential conservation laws
of the linear heat equation can be extended to this more general case.
More precisely, it is known that the space of characteristics of the local conservation laws of a linear partial differential equation~$Lu=0$
includes the functions of independent variables being  solutions of the adjoint equation~$L^*\lambda=0$.
Our conjecture is that any local or potential conservation law of a $(1+1)$-dimensional linear parabolic equation is equivalent to
one with the characteristic depending only on the time and space variables.
The conjecture was already tested for the linear heat equation~\cite{Popovych&Ivanova2004ConsLawsLanl}.
That is why, description of all possible potential symmetries of the linear heat equation will give a good hint
to solve the similar problem for arbitrary linear parabolic $(1+1)$-dimensional equation.
This will be the subject of our sequel paper.

\subsection*{Acknowledgements}
The authors are grateful to Prof. M.~Kunzinger for fruitful discussion.
Research of NMI was supported by the Erwin Schr\"odinger Institute for Mathematical Physics (Vienna, Austria) in form of Junior Fellowship
and by the grant of the President of Ukraine for young scientists (project number GP/F11/0061).
The research of ROP was supported by Austrian Science Fund (FWF), Lise Meitner project M923-N13.


\begin{thebibliography}{99}

\footnotesize

\bibitem{Akhatov&Gazizov&Ibragimov1987}
Akhatov~I.Sh., Gazizov~R.K. and Ibragimov~N.Kh.,
Group classification of equation of nonlinear filtration
{\it Dokl. AN SSSR}, 1987, V.293, 1033--1035.

\bibitem{Akhatov&Gazizov&Ibragimov1989}
Akhatov~I.Sh., Gazizov~R.K., Ibragimov~N.Kh.,
Nonlocal symmetries. A heuristic approach,
{\it Itogi Nauki i Tekhniki, Current problems in mathematics. Newest results}, 1989, V.34, 3--83
(Russian, translated in {\it J. Soviet Math.}, 1991, V.55, 1401--1450).

\bibitem{Anco&Bluman2002a}
Anco~S.C. and Bluman~G., Direct construction method for
conservation laws of partial differential equations.~I. Examples
of conservation law classifications, {\it Eur. J. Appl. Math.},
2002, V.13, Part 5, 545--566 (math-ph/0108023).

\bibitem{Anco&Bluman2002b}
Anco~S.C. and Bluman~G., Direct construction method for
conservation laws of partial differential equations.~II. General
treatment, {\it Eur. J. Appl. Math.}, 2002, V.13, Part 5, 567--585
(math-ph/0108024).

\bibitem{Bluman1989}
Bluman G.W. and Kumei S., {\it Symmetries and Differential Equations},
Springer-Verlag, New York, 1989.

\bibitem{Bluman&Reid&Kumei1988}
Bluman~G.W., Reid~G.J. and Kumei~S., New classes of symmetries for partial differential equations,
{\it J. Math. Phys.}, 1988, V.29, 806--811.

\bibitem{Dorodnitsyn&Svirshchevskii1983}
Dorodnitsyn~V.A. and Svirshchevskii~S.R., On Lie--B\"acklund
groups admitted by the heat equation with a~source, Preprint
N~101, Moscow, Keldysh Institute of Applied Mathematics of Academy
of Sciences USSR, 1983.

\bibitem{Ibragimov1985}
Ibragimov~N.H., Transformation groups applied to mathematical
physics, {\it Mathematics and its Applications (Soviet Series)},
Dordrecht, D. Reidel Publishing Co., 1985.

\bibitem{Ibragimov&Kara&Mahomed1998}
Ibragimov~N.H., Kara~A.H. and Mahomed~F.M.,
Lie--B\"acklund and Noether symmetries with applications,
{\it Nonlinear Dynam.}, 1998, V.15, N~2, 115--136.

\bibitem{Ivanova2004}
Ivanova N.,
Conservation laws and potential systems of diffusion-convection equations,
{\it Proceedings of Fifth International Conference ``Symmetry in Nonlinear Mathematical Physics''}
(23--29 June, 2003, Kyiv),
Kyiv, Institute of Mathematics, 2004, Part~1, 149--153 (math-ph/0404025).


\bibitem{Ivanova&Popovych&Sophocleous2005}
Ivanova N.M., Popovych R.O. and Sophocleous C.,
Conservation laws of variable coefficient diffusion-convection equations,
Proc. of 10th International Conference in Modern Group Analysis (MOGRAN X) (Larnaca, Cyprus, 2004),
2005, 108--115 (math-ph/0505015).

\bibitem{Kara&Mahomed2000}
Kara~A.H. and Mahomed~F.M.,
Relationship between symmetries and conservation laws,
{\it Internat. J. Theoret. Phys.}, 2000, V.39, N~1, 23--40.

\bibitem{Khamitova1982}
Khamitova R.S., The structure of a group and the basis of
conservation laws, {\it Teoret. Mat. Fiz.}, 1982, V.52, N~2,
244--251.

\bibitem{Mei&Zhang2006}
Mei J-q. and Zhang H.-q.,
Potential symmetries and associated conservation laws to Fokker--Planck and Burgers equation,
{\it Internat. J. Theoret. Phys.}, 2006, in press. 

\bibitem{Ovsiannikov1982}
Ovsiannikov~L.V., {\it Group analysis of differential equations},
New York, Academic Press, 1982.

\bibitem{Olver1986}
Olver P., {\it Applications of Lie groups to differential equations},
New York, Springer-Verlag, 1986.

\bibitem{Popovych&Ivanova2004ConsLawsLanl}
Popovych R.O. and Ivanova N.M., Hierarchy of conservation laws of
diffusion--convection equations, {\it J. Math. Phys.}, 2005, V.46,
043502, 22 p. (math-ph/0407008).

\bibitem{Popovych&Ivanova2003PETs}
Popovych R.O. and Ivanova N.M.,
Potential equivalence transformations for nonlinear diffusion--convection equations,
{\it J. Phys. A}, 2005, V.38, 3145--3155 (math-ph/0402066).

\bibitem{Pucci&Saccomandi1993}
Pucci E. and Saccomandi G.,
Potential symmetries and solutions by reduction of partial differential equations
{\it J. Phys. A}, 1993, V.26, 681--690.

\bibitem{Saccomandi1997}
Saccomandi G.,
Potential symmetries and direct reduction methods of order two,
{\it J. Phys. A}, 1997, V.30, 2211--2217.

\bibitem{Sophocleous1996}
Sophocleous~C.,
Potential symmetries of nonlinear diffusion-convection equations,
{\it J. Phys. A}, 1996, {V.29}, 6951--6959.

\bibitem{Steinberg&Wolf1981}
Steinberg~S. and Wolf~K.B.,
Symmetry, conserved quantities and moments in diffusive equations,
{\it J. Math. Anal. Appl.}, 1981, V.80, N~1, 36--45.

\bibitem{Wolf2002}
Wolf~T.
A comparison of four approaches to the calculation of conservation laws,
{\it Eur. J. Appl. Math.}, 2002, V.13, Part 5, 129--152.

\bibitem{Zharinov1986}
Zharinov~V.V., Conservation laws of evolution systems, {\it
Teoret. Mat. Fiz.}, 1986, V.68, N~2, 163--171.



\end{thebibliography}
\end{document}